\documentclass[graybox]{svmult}

\usepackage{mathptmx}       
\usepackage{helvet}         
\usepackage{courier}        
\usepackage{type1cm}        
\usepackage{amsmath}
\usepackage{makeidx}         
\usepackage{graphicx}        
\usepackage{multicol}        
\usepackage[bottom]{footmisc}
\usepackage{url}        
\usepackage{psfrag}        
\newcommand{\de}{{\rm \, d}}

\makeindex             

\begin{document}

\title*{Some Unusual Properties of Turbulent Convection and Dynamos in
  Rotating Spherical Shells} 
\titlerunning{Unusual Properties of Turbulent Convection and Dynamos} 
\author{F.~H.~Busse and R.~D.~Simitev}
\institute{F.~H.~Busse 
\at Institute of Physics, University of Bayreuth, D-95440 Bayreuth,
Germany, \email{busse@uni-bayreuth.de} 
\and R.~D.~Simitev 
\at Department of Mathematics, University of Glasgow, Glasgow G12 8QW,
UK} 

\maketitle

\abstract*{The dynamics of convecting fluids in rotating spherical
  shells is governed at Prandtl numbers of the order unity by the
  interaction between differential rotation and roll-like convection
  eddies. While the differential rotation is driven by the Reynolds
  stresses of the eddies, its shearing action inhibits convection and
  causes phenomena such as localized convection and turbulent
  relaxation oscillations. The response of the system is enriched in
  the case of dynamo action. Lorentz forces may brake either entirely
  or partially the geostrophic differential rotation and give rise to
  two rather different dynamo states. Bistability of turbulent dynamos
  exists for magnetic Prandtl numbers of the order unity.  While the
  ratios between mean magnetic and kinetic energies differ by a factor
  of 5 or more for the two dynamo states, the mean convective heat
  transports are nearly the same. They are much larger than in the
  absence of a magnetic field.} 

\section{Introduction}
\label{sec:1}
Convection of an electrically conducting fluid in a rotating system
represents a basic dynamical process in planetary interiors and in
stars. Astrophysicists and geophysicists have long been interested in
the mechanisms that govern the convective heat transport and the
generation of magnetic fields by convection in those systems. The
availability in recent years of large scale computer capacities has
permitted numerical simulations of detailed models for those
processes. Only the larger length scales can be taken into account in
those computations, of course, and eddy diffusivities are usually
introduced to model the influence of the smaller unresolved scales of
the turbulent flows.  

A difficulty arises from the fact that it can not generally be assumed
that all eddy diffusivities are equal. First, they apply to scalar as
well as to vector quantities, such as temperature and magnetic
fields. Secondly, the diffusivities in the absence of turbulence
differ enormously such that the turbulence may not be sufficiently
strong to equalize them. In the Earth's liquid core, for instance, the
magnetic diffusivity is large enough to be taken into account without
the consideration of an eddy contribution while a comparable eddy
viscosity would have to exceed the probable molecular viscosity value
by a factor of at least $10^6$. As has been demonstrated in the past
\cite{SB} the dynamics of convection in rotating spherical shells and
its dynamo action are very sensitive to ratios of diffusivities,
especially to the Prandtl number around its usually assumed value of
unity. 

More complex methods for treating effects of turbulence could
eventually be used, such as $k-\epsilon$-models, and the undoubtedly
important anisotropy of turbulent diffusivities in rotating systems
could also be considered. Since these influences are difficult to
evaluate, however, we shall restrict the attention in this paper to a
minimum of physical parameters. Inspite of this restriction a number
of characteristic features of convection and its dynamos in rotating
spherical shells can be demonstrated that are likely to exist as
coherent spatio-temporal structures in natural systems.  

\section{Mathematical formulation of the problem and methods of solution}

We consider a rotating spherical fluid shell of thickness $d$
and assume that a  static state exists with the temperature
distribution $T_S = T_0 - \beta d^2 r^2 /2$. Here $rd$ is the length
of the position vector with respect to the center of the sphere. 
The gravity field is given by $\vec g = - d \gamma \vec r$. In
addition to  $d$, the time $d^2 / \nu$,  the temperature $\nu^2 /
\gamma \alpha d^4$ and  the magnetic flux density $\nu ( \mu \varrho
)^{1/2} /d$ are used as scales for the dimensionless description of
the problem  where $\nu$ denotes the kinematic viscosity of the fluid,
$\kappa$ its thermal diffusivity, $\varrho$ its density and $\mu$ is
its magnetic permeability. Since we shall assume the Boussinesq
approximation material properties are regarded as constants except for
the temperature dependence of the density described by $\alpha \equiv
- ( \de \varrho/\de T)/\varrho$ which is taken into account only in
the gravity term. Both, the velocity field $\vec v$ and the magnetic
flux density $\vec B$, are solenoidal vector fields for which the
general representation 
\begin{gather}
\vec v = \nabla \times ( \nabla u \times \vec r) + \nabla w \times
\vec r, \qquad \qquad
\vec B = \nabla \times  ( \nabla h \times \vec r) + \nabla g \times
\vec r,
\end{gather}
can be employed. By multiplying the (curl)$^2$ and the curl of the
Navier-Stokes equations of motion by $\vec r$ we obtain two equations
for $u$ and $w$,   
\begin{gather}
\label{momentum}
[( \nabla^2 - \partial_t) {\cal L}_2 + \tau \partial_{\varphi} ]
\nabla^2 u +
\tau {\cal Q} w - {\cal L}_2 \Theta
= - \vec r \cdot \nabla \times [ \nabla \times ( \vec v \cdot
\nabla \vec v - \vec B \cdot \nabla \vec B)], \\
[( \nabla^2 - \partial_t) {\cal L}_2 + \tau \partial_{\varphi} ] w -
\tau {\cal Q}u
= \vec
r \cdot \nabla \times ( \vec v \cdot \nabla \vec v - \vec B \cdot
\nabla \vec B),
\end{gather}
where $\partial_t$ denotes the partial derivative with respect to time
$t$ and where $\partial_{\varphi}$ is the partial derivative with respect to
the angle $\varphi$ of a spherical system of coordinates $r, \theta,
\varphi$. For further details we refer to \cite{SB}. The operators ${\cal
L}_2$ and $\cal Q$ are defined by  
\begin{gather}
{\cal L}_2 \equiv - r^2 \nabla^2 + \partial_r ( r^2 \partial_r),
\nonumber\\
{\cal Q} \equiv r \cos \theta \nabla^2 - ({\cal L}_2 + r \partial_r )
( \cos \theta
\partial_r - r^{-1} \sin \theta \partial_{\theta}). \nonumber
\end{gather}
The heat equation for the dimensionless deviation $\Theta$ from the
static temperature distribution can be written in the form
\begin{equation}
\label{heat}
\nabla^2 \Theta + R{\cal L}_2 u = P ( \partial_t + \vec v \cdot \nabla ) \Theta,
\end{equation}
and the equations for $h$ and $g$ are obtained through the multiplication of the 
equation of induction and of its curl by $\vec r$,
\begin{gather}
\label{induction}
\nabla^2 {\cal L}_2 h = P_m [ \partial_t {\cal L}_2 h - \vec r \cdot
\nabla \times ( \vec v \times \vec B )], \\
\nabla^2 {\cal L}_2 g = P_m [ \partial_t {\cal L}_2 g - \vec r \cdot
\nabla \times ( \nabla \times ( \vec v \times \vec B ))].
\end{gather}
The Rayleigh number $R$,
the Coriolis number $\tau$, the Prandtl number $P$ and the magnetic
Prandtl number $P_m$ are defined by 
\begin{equation}
R = \frac{\alpha \gamma \beta d^6}{\nu \kappa} , 
\qquad \tau = \frac{2
\Omega d^2}{\nu} , \qquad P = \frac{\nu}{\kappa} , \qquad P_m = \frac{\nu}{\lambda},
\end{equation}
where $\lambda$ is the magnetic diffusivity. For the static
temperature distribution we have chosen the case of a homogeneously
heated sphere. This state is traditionally used for the analysis of
convection in self-gravitating spheres and offers the numerical
advantage that for Rayleigh numbers close to the critical value $R_c$
the strength of convection does not differ much near the inner and
outer boundaries. As the Rayleigh number increases beyond $R_c$ heat
enters increasingly at the inner boundary and is delivered by
convection to the outer boundary. When $R$ reaches a high multiple of
$R_c$ the heat generated internally in the fluid becomes negligible in
comparison to the heat transported by convection from the inner to the outer boundary.
\begin{figure}[t]
\begin{center}
\includegraphics[bb=340 636 456 750,clip,width=4.5cm]{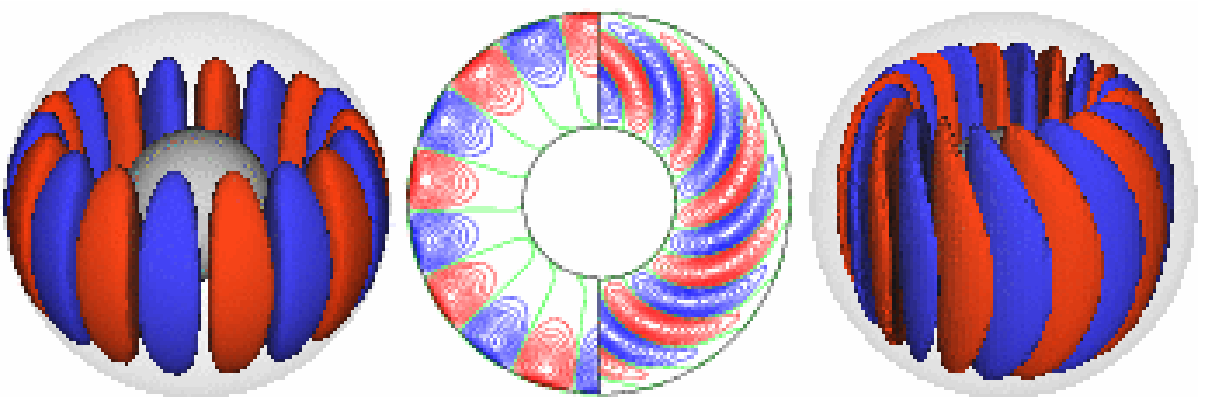}
\caption{Convection columns in a rotating spherical fluid shell for
  $\tau=10^4, R=2.8\cdot 10^5, P=1$. Blue and red surfaces
  correspond to a constant positive and negative value of the radial
  velocity.}
\end{center}
\end{figure}

Fixed temperatures and stress-free boundaries, 
\begin{equation}
\label{vbc}
u = \partial^2_{rr}u = \partial_r (w/r) = \Theta=0 
\quad \mbox{ at } r=r_i \equiv \eta/(1-\eta) \mbox{ and } r=r_o \equiv 1/(1-\eta),
\end{equation}
will be assumed where $\eta$ denotes radius ratio,
$\eta=r_i/r_o$. In the following only $\eta=0.4$ will be used.
For the magnetic field electrically insulating boundaries are assumed
such that the poloidal function $h$ must be  matched to the function
$\hat h$ which describes the   potential fields outside the fluid
shell,   
\begin{equation}
\label{mbc}
g = h-\hat h = \partial_r ( h-\hat h)=0 
\quad \mbox{ at } r=r_i \equiv \eta/(1-\eta) \mbox{ and } r=r_o \equiv 1/(1-\eta).
\end{equation}
But computations for the case of an inner boundary with no-slip
conditions and an electrical conductivity equal to that of the fluid
have also been done \cite{SB}. The numerical integration of the equations
 together with the boundary
conditions proceeds with the pseudo-spectral  
method as described in \cite{TB7} and \cite{T99} 
which is based on an expansion of all dependent variables in
spherical harmonics for the $\theta , \varphi$-dependences, i.e. 
\begin{equation}
u = \sum \limits_{l,m} U_l^m (r,t) P_l^m ( \cos \theta ) \exp \{ im \varphi \}
\end{equation}
and analogous expressions for the other variables, $w, \Theta, h$ and $g$. 
$P_l^m$ denotes the associated Legendre functions.
For the $r$-dependence expansions in Chebychev polynomials are used. 
For further details see also \cite{SB}.
For the non-magnetic convection calculations to be reported in the
following a minimum of 33 collocation points in the radial direction
and spherical harmonics up to the order 64 have been used. The
resolution has been increased to 41 or 55 collocation points and
spherical harmonics up to the order 96 or 128 in the case of dynamo
simulations. 
\begin{figure}[t]
\begin{center}
\includegraphics[width=12cm,height=8cm]{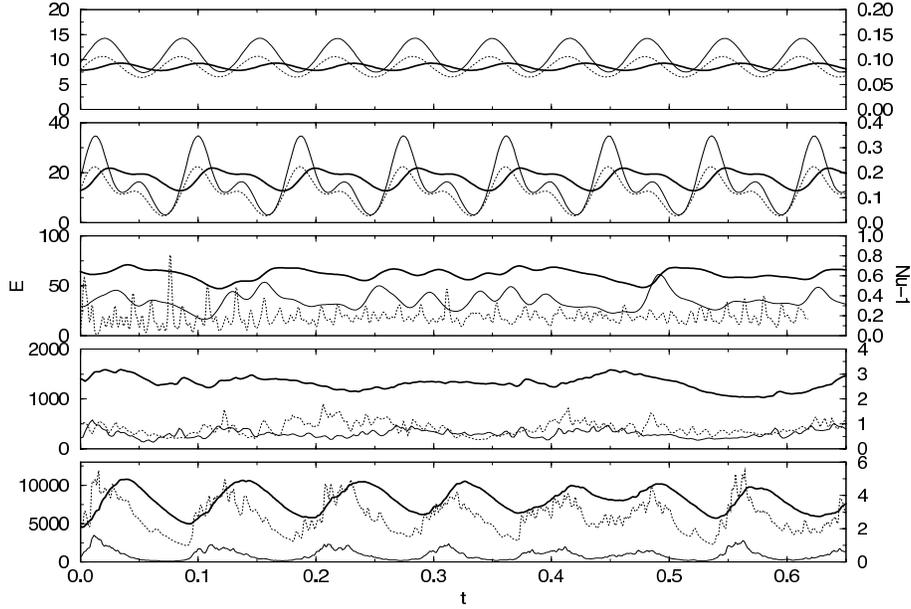}
\caption{Time series of energy densities of convection for $P = 1$,
  $\tau= 10^4$ and $R = 2.8\times10^5$ ,  
$3.0 \times 10^5$, $3.5 \times 10^5$, $7 \times 10^5$, $12 \times
10^5$, (from top to bottom). Solid and dashed lines
indicate $\overline{E}_t$ and $\widetilde{E}_t$,
respectively. The Nusselt number $Nu_i$ is indicated by dotted lines and measured at the right ordinate.}
\end{center} 
\end{figure}

\section{Convection in rotating spherical shells}

For an introduction to the problem of convection in spherical shells
we refer to the review \cite{Bu} and to the respective chapter in the
book \cite {DS}. Additional information can be found in the papers by
Grote and Busse \cite{GB}, Jones et al. \cite{J}, Christensen
\cite{C02} and Simitev and Busse \cite{SB1}. Typically the onset of
convection occurs in the form of progradely propagating thermal Rossby
waves as illustrated in figure 1. Only for low Prandtl numbers $P$,
i.e. $P < 10/\sqrt{\tau}$ according to \cite{A97}, the onset occurs in
the form of inertial waves attached to the outer equatorial boundary
of the fluid shell as is discussed in \cite{ZB},\cite{Z94},\cite{A97}.   

Because of the symmetry of the velocity field with respect to the
equatorial plane it is sufficient 
to plot streamlines in this plane, given by $r \partial u/\partial r=
const.$, to characterize the convection flow. This has been
done in figures 3 and 5. Even in the case of turbulent convection
the part of the velocity field that is antisymmetric with respect to
the equatorial plane is rather small as long as the parameter $\tau$ is sufficiently large. 

As the Rayleigh number $R$ grows beyond its critical value $R_c$,
the thermal Rossby waves become modified by a
sequence of bifurcations similar to those found in
other problems of convection. First, oscillations of the amplitude
are observed, then another bifurcation causes a
low wavenumber modulation as function of the azimuth
\cite{SB1}. Finally, a chaotic state of convection is obtained.   
\begin{figure}[t] 
\begin{center}
\includegraphics[width=9cm]{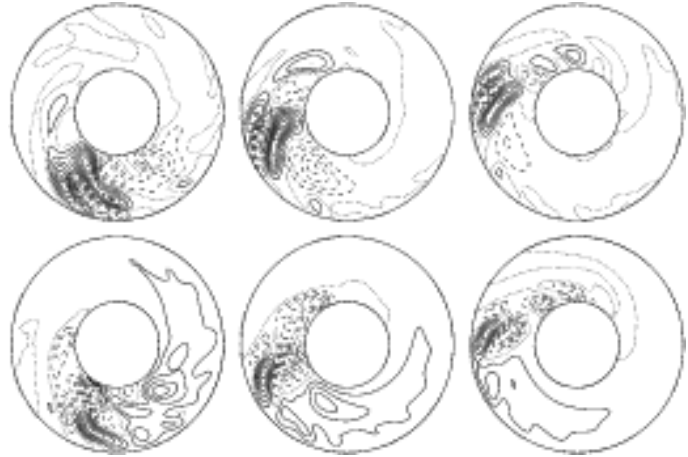}
\caption{Localized convection for $R=7 \times 10^5, \tau =1.5 \times 10^4, P=0.5$
The streamlines, $r \partial u / \partial \varphi =$ const. (upper row) and the
isotherms, $\Theta =$ const. (lower row), are shown in the equatorial plane for
equidistant times (from left to right) with $\Delta t = 0.03$.}
\end{center} 
\end{figure}

\section{Chaotic Convection}

The sequence of transitions can also visualized through the time
dependence of average quantities such as the  contributions to the
kinetic energy density. These are defined by 
\begin{subequations}
\begin{align}
\label{7}
\bar E_p = \langle \mid \nabla \times ( \nabla \bar u \times \vec r )
\mid^2 \rangle /2 , \quad
& \bar E_t = \langle \mid \nabla \bar w \times \vec r \mid^2 \rangle/2
\\
\label{8}
\widetilde E_p =  \langle \mid \nabla \times ( \nabla \widetilde u
\times \vec r)
\mid^2 \rangle /2, \quad&
\widetilde E_t = \langle \mid \nabla \widetilde w \times\vec r \mid^2
\rangle/2
\end{align}
\end{subequations}
where the angular brackets indicate the average over the fluid shell 
and where $\bar u$ refers to the azimuthally averaged 
component of $u$ and $\widetilde u$ is given by $\widetilde u = u -
\bar u $. At low supercritical Rayleigh numbers the energy densities
corresponding to steadily drifting thermal Rossby waves are constant
in time and are not included in figure 2. The onset of vacillations
manifests itself in the sinusoidal oscillations of the kinetic
energies as shown in the top plot of figure 2.  
Also plotted in figure 2 is the Nusselt number $Nu_i$  measuring the
efficiency of the convective heat transport at the inner boundary,  
\begin{equation}
Nu_i=1- \frac{P}{r_i} \left.\frac{d \overline{\overline{\Theta}}}{d r}\right|_{r=r_i}
\end{equation}  
where the double bar indicates the average over the spherical surface. 
$\bar E_t$ describes the energy density of the differential rotation
which increases strongly with increasing $R$ as can be noticed in the lower plots of
figure 2. This increase is caused by the strong azimuthal Reynolds stress exerted by the
convection eddies resulting from their inclination with respect to
radial direction as is apparent in  figure 1. The increasing shear
of the differential rotation tends to inhibit convection, however, in
that it shears off the convection eddies. This is a consequence of the
nearly two-dimensional nature of the dynamics in a rotating system:
Because of the requirement that the structure of convection approaches
closely the Taylor-Proudman condition, there is no possibility for a
reorientation of the convection rolls as happens in 
non-rotating systems. In the rotating sphere convection thus generates
the agent that tends to destroy it. A precarious 
balance in the form localized convection is the result. As shown in
figure 3 convection occurs only in a restricted azimuthal section of
the spherical shell where its amplitude is strong enough to
overcome the inhibiting influence of the shear. For the geostrophic
zonal flow it does not matter whether it is driven locally or more uniformly around
the azimuth. Since in the non-convecting region thermal buoyancy
accumulates, the advection of this buoyancy by the differential
rotation  strengthens and stabilizes the localized convection.  

At even higher Rayleigh numbers this balance no longer works and
instead of a  localization in space the localization of convection in
time is initiated as is 
shown in figures 4 and 5. Here convection exist only for a
short period while the differential rotation is sufficiently weak. As
the amplitude of convection grows, the differential rotation grows even
more strongly since the Reynolds stress increases with the square of the
amplitude. Soon the shearing action becomes strong enough to cut off
convection. Now a viscous diffusion time must pass before the
differential rotation has decayed sufficiently such that convection
may start growing again. It is remarkable to see how the chaotic
system exhibits its nearly periodic relaxation oscillations as shown
in figure 4. 
\begin{figure}[t] 
\begin{center}
\includegraphics[height=10cm,angle=-90]{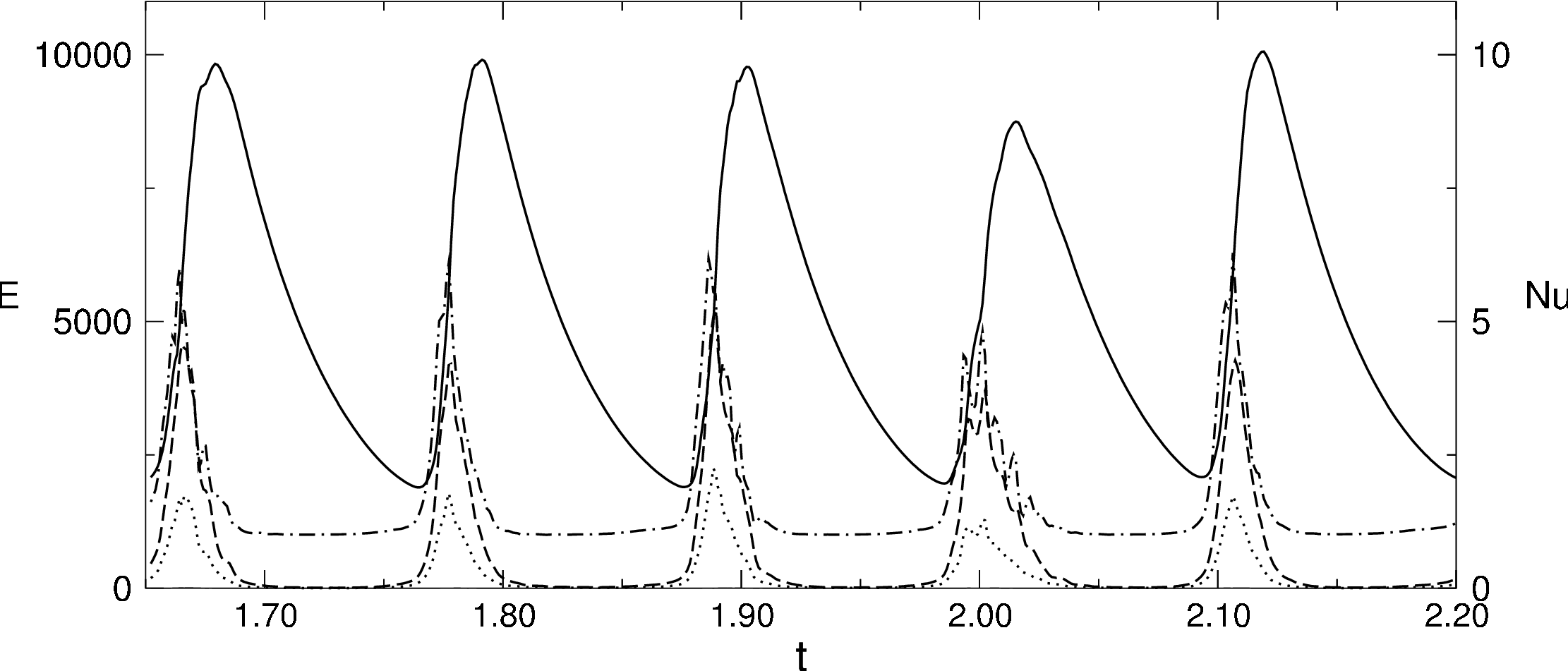}
\caption{Relaxation oscillations of chaotic convection  in the case
  $\tau=10^4, R=6.5\cdot 10^5, P=0.5$. The energy densities $\overline{E_t}$
  (solid line), $\widetilde E_t$ (dashed line), $\widetilde E_p$ (dotted line) and the
  Nusselt number (dash-dotted line, right ordinate) are shown as function of time.} 
\end{center} 
\end{figure}

In figure 5 a sequence of plots is shown at four instances around the
time of a convection peak. At first there is hardly any convection, -
the dotted lines just indicate zero. At the next instance the
differential rotation as shown by the upper row has decayed
sufficiently such that convection columns can grow reaching nearly
their maximum amplitude in the third plot. At the same time the
differential rotation has grown as well and begins to exert its
inhibiting effect such that convection decays at the fourth instance
of the sequence, while the differential rotation reaches its
maximum. It should be mentioned that localized convection and
relaxation oscillations occur at moderate Prandtl numbers of the order
unity or less. At higher values of $P$ Reynolds stresses are no longer
sufficiently powerful to generate a strong differential
rotation. Instead variations of the temperature field caused by
the dependence of the convective heat transport on latitude induce a
differential rotation in the form of a thermal wind.  
\begin{figure}[t] 
\begin{center}
\includegraphics[width=12cm]{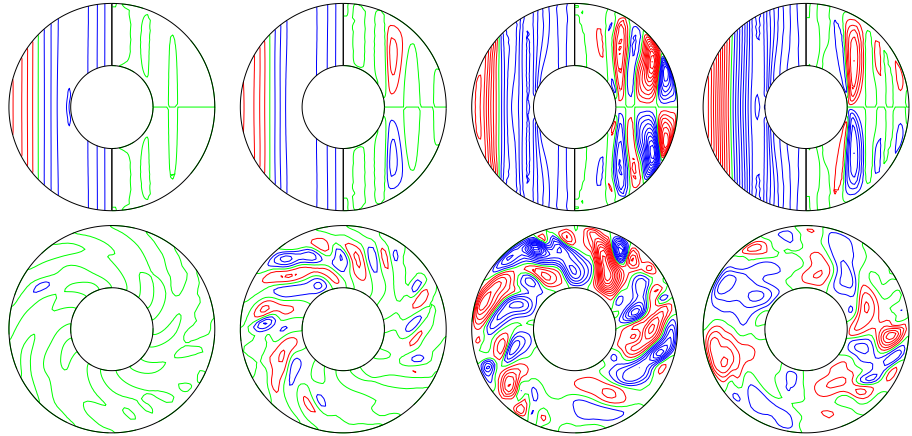}
\caption{Sequence of plots starting at $t=2.31143$ and equidistant in
  time $(\Delta t=0.01$) for 
  the same case as in Fig. 8.   Lines of constant
  $\bar{u}_{\varphi}$ and 
streamlines $r \sin\theta\partial_\theta \overline{h}=$const.
 in the meridional plane, are shown in  the left 
  and right halves, respectively, of the upper row. The lower 
  row shows corresponding streamlines, $r \partial u / \partial \varphi =$
  const., in the equatorial plane.}
\end{center} 
\end{figure}
\begin{figure} 
\begin{center}
\includegraphics[height=7cm,angle=-90]{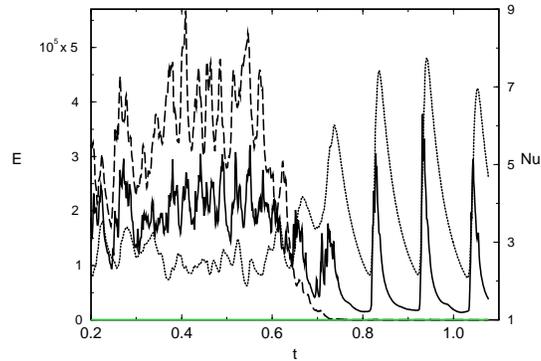}
\caption{Transition from a dynamo state to a state of chaotic
  relaxation oscillations for $\tau=1.5\times10^4, R=1.2\times 10^6,  P=
P_m=0.5$. The energy density $\overline{E}_t$ (dotted line),
  the total magnetic energy density  (dashed line)
  and the Nusselt number  $Nu_i$ (solid line, right ordinate) are shown as function of
  time.}  
\end{center} 
\end{figure}

The convective heat transport in the case of localized convection as
well as in the case of relaxation oscillations is much reduced, of
course, relative to a case without strong differential rotation. This
causes the magnetic field to enter the
problem in a crucial way provided the electrical conductivity is sufficiently high. By putting  brakes on the differential
rotation through its Lorentz force the magnetic field permits a much
higher heat transport than would be possible in an
electrically-insulating fluid. This is the basic reason that the
Earth's core as well as other  planets with convecting cores and
rotating stars exhibit magnetic fields. A demonstration of this effect
is seen in figure 6 where by chance the convection driven dynamo was
just marginal such that it could not recover after a downward
fluctuation of the magnetic field. Hence the relaxation oscillations
with their much reduced average heat flux take over from the dynamo
state.   

\section{Distinct turbulent dynamos at identical parameter values}

Convection driven dynamos in rotating spherical fluid shells are often
subcritical as is apparent in figure 6, for instance. At somewhat
higher Rayleigh numbers convection with a strong magnetic field
will persist. On the other hand the dynamo will decay when the magnetic field is
artificially reduced to, say, a quarter of its averaged energy. There
thus exists the possibility of a convection driven
dynamo state and of a non-magnetic convection state at identical
values of the external parameters $R, \tau, P$. The bistable
coexistence of a non-magnetic convection state and a dynamo state is
typical for subcritical bifurcations as in the analogous coexistence
of laminar   and turbulent states in shear flows.  
\begin{figure}[t]
\begin{center}
\includegraphics[height=8cm]{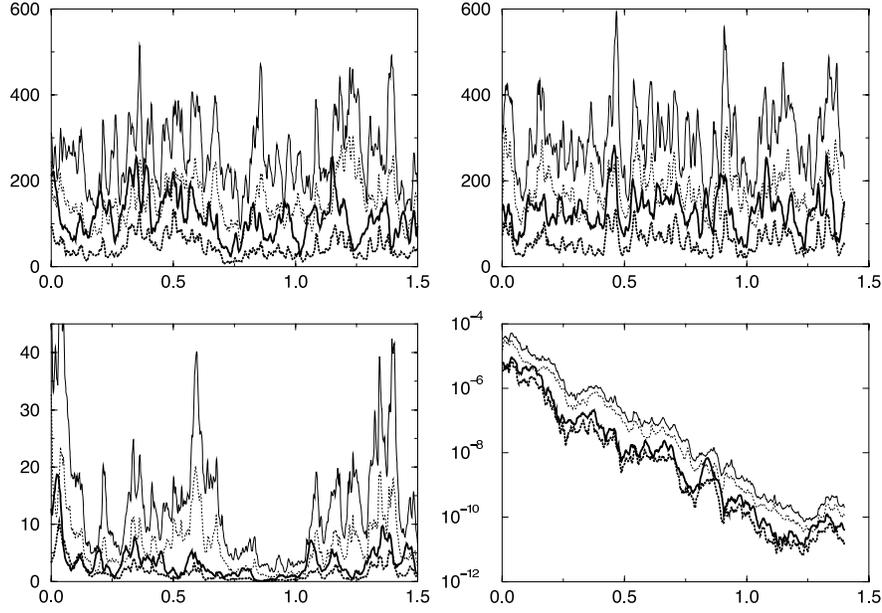}
\caption{Two distinct dynamos at identical parameter values,
  $\tau=5\times10^3, R=5\times10^5, P=P_m=1$. Upper (lower) plots show
  time series of quadrupolar (dipolar) magnetic energy
  densities. Thick lines indicate mean torodial (solid lines) and
  poloidal (dashed lines) energy components. Thin lines indicate the
  same for the fluctuating components.} 
\end{center} 
\end{figure}
\begin{figure} 
\begin{center}
\hspace*{-5mm}
\includegraphics[width=12cm,clip=]{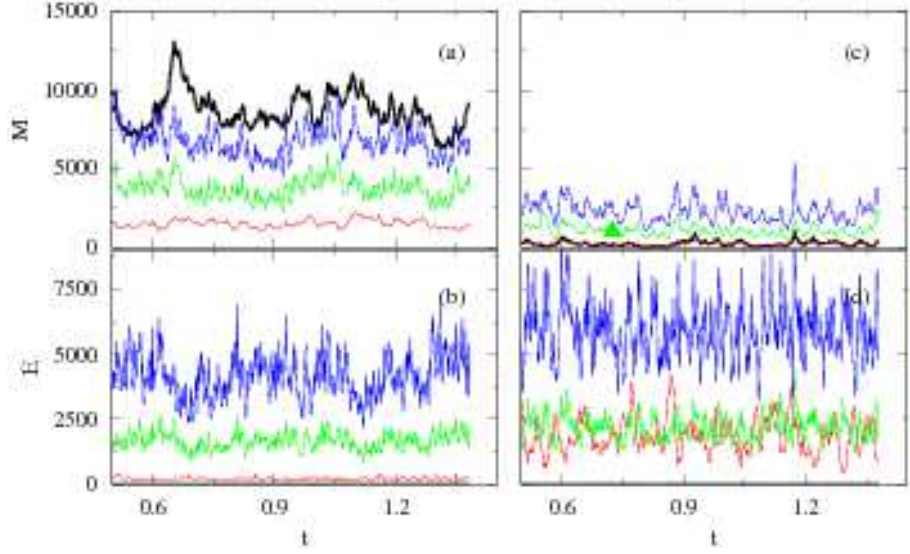}
\caption[]
{ Time series of two different chaotic attractors are
  shown - a MD (left column (a,b)) and a
FD dynamo (right column (c,d)) both in the case $R=3.5\times10^6$,
$\tau=4\times10^4$,
  $P=0.5$ and  $P_m=1$.
The top two panels (a,c) show magnetic energy densities. and the
bottom two panels (b,d) show kinetic energy densities in the presence
of the magnetic field. The components $\overline{M}_p$ are shown by
thick solid black lines, while
$\overline{X}_t$, $\widetilde {X}_p$, and $\widetilde{X}_t$ are 
shown in red green and blue
respectively. 
$X$ stands for either $M$ or $E$.
}
\end{center} 
\end{figure}
\begin{figure} 
\begin{center}
\hspace*{-5mm}
\includegraphics[width=12cm,clip=]{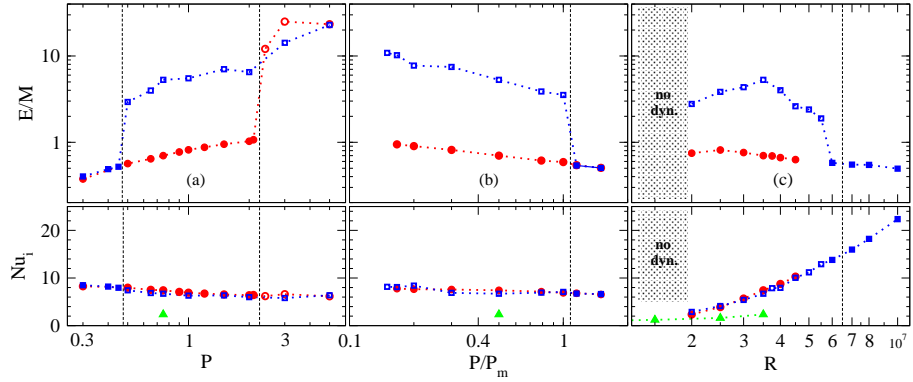}
\caption{The upper row shows the hysteresis effect in
  the ratio of magnetic to kinetic energy, $E/M$, at $\tau=3\times10^4$
  (a) as a function of the Prandtl number in the case
  of $R=3.5\times10^6$, $P/P_m=0.5$; (b) as a function of
  the ratio  $P/P_m$ in the case of $R=3.5\times10^6$,
  $P=0.75$ and (c)  as a function of the Rayleigh number in the case
  $P=0.75$, $P_m=1.5$. Full and empty 
   symbols indicate FD and   MD dynamos, respectively and circles and
   squares indicate the two hysteresis branches.
 The critical value of $R$ for the onset of
  thermal convection for the cases shown in (c) is
  $R_c=659145$. A   transition from FD to MD dynamos as $P/P_m$
  decreases in   (b) is expected, but is not indicated owing to lack
  of data. The lower row shows the value $Nu_i$ of the Nusselt number
  at   $r=r_i$ for the same dynamo cases. Values for non-magnetic
  convection are  indicated by triangles for comparison. }
\end{center} 
\end{figure}

More surprising is the fact that two different turbulent dynamo states
can exist at identical values of the external parameters which now should
include the magnetic Prandtl number  $P_m$. An example is shown in
figure 7 where a spherical dynamo of mixed parity, i.e. with dipolar
and quadrupolar components, and a purely quadrupolar dynamo evolve in
time at identical external parameters. In the latter case the
exponential decay of dipolar disturbances is clearly demonstrated. In
figure 7 magnetic energy densities have been plotted that are defined
in  analogy to expressions (12), 
\begin{subequations}
\begin{align}
\overline{M}_p = \langle \mid \nabla \times ( \nabla \bar h \times
\vec r
) \mid^2 \rangle /2 , \quad &
\overline{M}_t = \langle \mid \nabla \bar g_t \times \vec r \mid^2
\rangle/2 \\
\widetilde{M}_p = \langle \mid \nabla \times ( \nabla \widetilde h_p
\times \vec r)
\mid^2 \rangle/2 , \quad&
\widetilde{M}_t =  \langle \mid \nabla \widetilde g_t \times
\vec r \mid^2 \rangle /2.
\end{align}
\end{subequations}

An even more surprising case is that of two convection driven dynamos
without any distinction in symmetry, just with differences in the
magnitude of various energy densities  as is apparent from figure
8. While a strong mean poloidal magnetic field as shown in the left
half of figure 8 acts as an efficient brake on the differential
rotation as measured by $\overline{E}_t$, it also inhibits
convection. The alternative dynamo on the right side of the figure is
characterized by a relatively weak mean magnetic field and dominant
fluctuating components. Here the kinetic energy densities of
convection are larger, but the differential rotation is still much
weaker than it would be in the non-magnetic case.  

The bistability in the form of two different types of dynamos is not a
singular phenomenon, but exists over an extended region of the
parameter space. Since the region includes values of the Prandtl
number of the order unity which has been the preferred value of $P$ in
many simulations of convection driven dynamos the phenomenon of
bistable dynamos is of considerable importance. The first results of
our extensive numerical simulations can be found in \cite{SB9}. A typical
diagram is shown in figure 9. The extended regime of coexistence of
the two types of dynamo is bounded by transitions where one of the two
dynamos ceases to be stable and evolves into the other one within a
magnetic diffusion time.  A basic reason for the competitiveness
of both dynamos is that they exhibit essentially  the same convective
heat transport as measured by the Nusselt number $Nu_i$.   The lower
half of figure 9 not only demonstrates the surprising coincidence of
the heat transports of the two dynamo types, but also indicates that these
heat transports by far exceed those found in the absence of a magnetic
field.  

Recently  the computations have been extended to higher rotation rates
as shown in figure 10. At $\tau=4\times10^4$ the same phenomenon of
bistability has been found as at $\tau=3\times10^4$. Again, the two
kinds of dynamo exhibit the same heat transport as indicated in the
lower part of figure 10. 
\begin{figure}[t] 
\begin{center}
\psfrag{Nui}{$Nu_i$}
\psfrag{MpfMpm }{$\widetilde{M}_p/\overline{M}_p$}
\includegraphics[width=7cm,clip=]{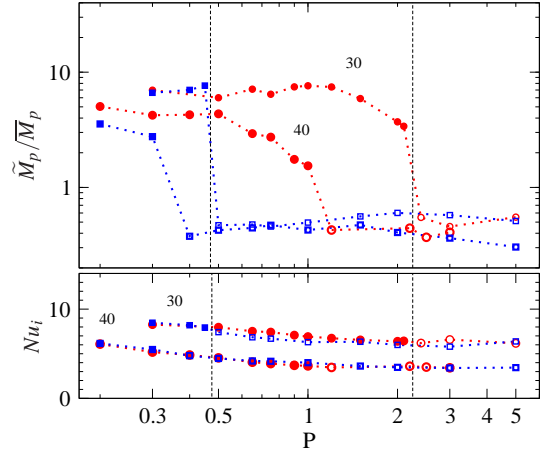}
\caption{The upper part shows the ratio between the energy densities
  of the fluctuating and the mean poloidal  magnetic field,
  $\widetilde{M}_p/\overline{M}_p$, as a function of the Prandtl number at
  $\tau=3\times10^4$ (denoted by 30 in plot)
  and at  $\tau=4\times10^4$
  (denoted by 40 in plot) by symbols analogous to those in fig 10. 
  In all cases $R=3.5\times10^6$ and $P_m=2 P$. 
  The lower part shows the value $Nu_i$ of the Nusselt number for the same dynamo cases.} 
\end{center} 
\end{figure}

\section{Concluding Remarks}

The existence of two distinct turbulent states is a rare phenomenon,
although examples exist in non-magnetic hydrodynamics, see, for
instance, \cite{Ra} \cite{Mu} . In magnetohydrodynamics an
electrically conducting fluid in the presence of a magnetic field
offers new degrees of freedom which allow more than a single balance
between the various forces operating in  turbulent states. Initial
conditions thus determine which of the competing states is actually
realized.    

The possibility of bistability could be of interest for the
interpretation of planetary and stellar magnetism. Magnetic hysteresis
effects associated with stellar oscillations may eventually be
explained in this way. Anyone involved with numerical simulations of
convection driven dynamos should be aware that his solutions could
change drastically after different  initial conditions have been
introduced.   



%
\begin{figure}[b]
\end{figure}

\end{document}